\documentclass[twocolumn,prb,showpacs]{revtex4}
\usepackage{fancyhdr}
\usepackage{amsmath}
\usepackage{subfigure}
\usepackage{epsfig}
\usepackage{amssymb}
\usepackage{units} 
\usepackage{pifont}
\usepackage{textcomp}
\usepackage{gensymb}
\usepackage{pifont}
\usepackage{enumerate}
\usepackage{ulem}
\usepackage{color}
\begin{document}

\title{Low energy bands and transport properties of chromium arsenide}

\author{Carmine Autieri}

\affiliation{CNR-SPIN, I-67100 Coppito, (L'Aquila), Italy}

\author{Giuseppe Cuono}

\affiliation{Dipartimento di Fisica ``E. R. Caianiello'', Universit\`a di
	Salerno, I-84084 Fisciano (Salerno), Italy}

\author{Filomena Forte}

\affiliation{Dipartimento di Fisica ``E. R. Caianiello'', Universit\`a di
	Salerno, I-84084 Fisciano (Salerno), Italy}

\affiliation{CNR-SPIN, I-84084 Fisciano (Salerno), Italy}

\author{Canio Noce}

\affiliation{Dipartimento di Fisica ``E. R. Caianiello'', Universit\`a di
	Salerno, I-84084 Fisciano (Salerno), Italy}

\affiliation{CNR-SPIN, I-84084 Fisciano (Salerno), Italy}

\date{\today}
\begin{abstract}

We apply a method that combines the tight-binding approximation and the L\"{o}wdin down-folding procedure to evaluate the electronic band structure of the newly discovered pressure-induced superconductor CrAs. By integrating out all low-lying arsenic degrees of freedom, we derive an effective Hamiltonian model describing the Cr $d$ bands near the Fermi level. We calculate and make predictions for the energy spectra, the Fermi surface, the density of states and transport and magnetic properties of this compound. Our results are consistent with local-density approximation calculations as well as they show good agreement with available experimental data for resistivity and Cr magnetic moment.

\end{abstract}


\maketitle

\section{Introduction}

In the recent years superconducting materials such as heavy-fermion compounds,~\cite{movshovich01} high transition-temperature cuprate superconductors,~\cite{vanharlingen95} strontium ruthenate superconductor~\cite{mackenzie03} and iron-pnictide superconductors~\cite{mazin08} have been extensively investigated due to their unconventional properties.~\cite{goll06} A common feature of these materials is that superconductivity appears due to the instability of some degree of freedom, which is mainly induced on the verge of a magnetically ordered phase driven by suitable external tuning parameters.\\
Very recently, pressure-induced superconductivity was discovered in CrAs in the vicinity of the helimagnetic (HM) phase.~\cite{wu14,kotegawa14}
This is the first example of superconductivity found in a Cr-based magnetic system. Previous measurements have shown that at ambient pressure CrAs undergoes a first- order phase transition to a non-collinear HM at $T_{N}\sim$ 265 K, ~\cite{wu10,wu14} where the propagation vector is found to be parallel to the $c$ axis and the magnetic moments lie in the $ab$ plane. Recent resistivity measurements under pressure revealed that the magnetic ordering temperature $T_{N}$ drastically decreases with pressure and that the magnetic order is completely suppressed above a critical pressure $P_c\sim$ 0.7 GPa. ~\cite{wu14,kotegawa15} Remarkably, superconductivity was discovered to appear on suppression of the magnetic phase, displaying a maximum superconducting transition temperature $T_c\sim$ 2.2 K at about 1 GPa. Increasing the pressure further decreases $T_c$, and the superconducting phase adopts a dome-like shape.~\cite{wu14,varma99,noce00,noce02,vandermarel03,jiang09,shibauchi14,seo15} This striking analogy to many superconducting systems suggested a possible unconventional pairing mechanism where the critical spin fluctuations could act as the glue medium for Cooper pairing.~\cite{wu14}\\
Wu $et\ al.$ reported the onset of superconductivity already at ∼0.3 GPa and a gradual increase of the superconducting volume fraction up to $P_c$. Accurate muon spin rotation measurements performed on CrAs powder samples, also revealed the existence of a region of coexistence in the intermediate pressure region, where the superconducting and the magnetic volume fractions are spatially phase separated and compete each others. ~\cite{khasanov15}
Moreover, a very recent nuclear quadrupole resonance study under pressure reported that the internal field in the helimagnetic state only decreases slowly with increasing pressure, but maintains a large value close to $P_c$.~\cite{kotegawa15} This indicates that the pressure-induced suppression of the magnetic order is of the first order. Therefore, even though substantial fluctuations are present in the paramagnetic state, the system is not close to quantum criticality. ~\cite{kotegawa15} It has also been revealed that the nuclear spin-lattice relaxation rate in CrAs shows substantial magnetic fluctuations, but does not display a coherence peak in the superconducting state, indicating an unconventional pairing mechanism.~\cite{kotegawa15} On the contrary, the phase separation scenario between magnetism and superconductivity together with the observation that the superfluid density $\rho_s$ scales with the critical temperature as $\sim T_c ^{3.2}$ have been interpreted as indicative of a conventional mechanism of pairing in CrAs.~\cite{khasanov15}\\
Concerning the normal phase properties at ambient pressure, a $T^2$ dependence of resistivity is observed at low temperatures, supporting a Fermi-liquid behaviour. The Kadowaki-Woods ratio is found to be $1\times10^{-5}\mu \Omega$ cm mol$^2$ K$^2$ mJ$^{-2}$, which fits well to the universal value of many heavy fermion compounds.~\cite{kadowaki86}  The first order magnetic transition at $T_{N}$ manifests itself via sharp changes of resistivity and susceptibility. ~\cite{} Above this temperature, a linear-temperature dependence of the magnetic susceptibility is observed up to $\sim700$~K.~\cite{wu10} Moreover, it has been shown that the first order magnetic transition at $T_{N}$ is accompanied abrupt changes of the lattice parameters, while a lowering of the crystal structure symmetry has not been reported.~\cite{} Neutron diffraction measurements~\cite{keller15,shen16}established a substantial  Cr moment of  1.7 $\mu_B$ lying essentially within the $ab$ plane.~\cite{} 
Such an interplay between structural, magnetic, and electronic properties at $T_{N}$ is well established in many transition metal compounds and is also expected to be crucial in CrAs in making the external pressure a very effective tool in fine tuning its ground-state. ~\cite{Cuoco06,Cuoco10} 

From a theoretical point of view, only ab-initio calculations based on first-principles density functional theory have been reported so far.~\cite{noce16}. The numerical simulations show that the magnetic and electronic properties strongly depend on the size of the unit cell, and the predicted physical quantities, including the Cr magnetic moment, are in good agreement with the experimental data.

In this paper we present the electronic structure and the magnetic properties in the normal state of the CrAs, investigated through the application of the tight-binding method.
When questioning about the opportunity to adopt such a single-particle description that entirely neglects electronic correlations, one should consider the following remarks.
It is known that materials whose resistances exceed the  Mott-Ioffe-Regel limit are known as bad metals,\cite{emery95} and this property is ubiquitous feature of the normal state of strongly correlated  materials.~\cite{hussey04} On the contrary, the high-temperature resistivity of CrAs indicates a saturation according to the Mott-Ioffe-Regel behaviour.~\cite{luo17} The latter consideration, together with the satisfying prediction for the local magnetic moment as inferred by the ab-initio calculations above mentioned, suggest that the CrAs may indeed be considered as a weakly correlated material.~\cite{noce16} This implies that the band structure of CrAs as obtained from local-density approximation calculations as well as from the tight-binding approach may give significant insights on the effective band spectra of this material.

In particular, in this paper we will adopt a modified tight-binding approach that combines the tight-binding approximation and the L\"{o}wdin down-folding technique.~\cite{lowdin50} In our procedure, we first derive a tight-binding model based on the Wannier transformation of the ab-initio results.~\cite{noce16} Subsequently, the 'folding'  procedure allows to replace the problem of the diagonalization of the complete full-range tight-binding Hamiltonian with that of an auxiliary matrix whose rank is definitely lower. L\"{o}wdin's technique  has  been successfully applied to e.g. cuprate superconductors,~\cite{andersen95} as well as to strontium ruthenate superconductor~\cite{noce99}, and it is particularly useful every time one is concerned only with a limited range of energy, e.g. few eV around the Fermi level.
We would like to point out that such a procedure is completely consistent with the full ab-initio calculations, since its parameters are the ab-initio derived overlap integrals for orbitals and the matrix elements. On the other hand, it avoids the complication of the conventional ab-initio calculations in cases where the unit cell contains many atoms, and allows to set up an effective Hamiltonian projected on the Cr electronic degrees of freedom and to obtain the analytical expressions for the low energy bands which are ready to be used to evaluate relevant physical quantities. In particular, we have been able to calculate the electronic band structure, the  corresponding density  of  states and the Fermi surface of CrAs, together with transport and magnetic properties of this compound. Our results are consistent with local-density approximation calculations and show good agreement with available experimental data for resistivity and Cr magnetic moment.

The paper is organized as follows: in the next section we will introduce the model Hamiltonian describing the CrAs and present the method adopted to diagonalise the reduced L\"{o}wdin Hamiltonian; in this way we will determine the electronic properties of CrAs looking at the energy spectrum, the Fermi surface and at the density of states; in Sec. III we will show the results for the electric resistivity obtained by means of the Boltzmann equation, and the  magnetic properties calculated within a self-consistent approach. The last section contains the conclusions and some remarks.

\section{Electronic Properties}

\begin{figure}
	\centering
	\includegraphics[width=8cm,angle=0]{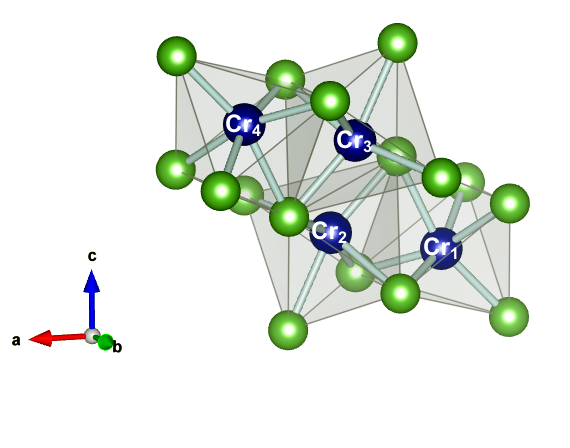}
	\caption{Crystal structure of the CrAs. Cr and As are shown as blue and green spheres, respectively. All the Cr atoms are equivalent even though the Cr-Cr distances are different.}
	\label{fig:CrAs}
\end{figure}

CrAs belongs to the family of transition-metal pnictides with the general formula AB (A=transition metal, B=P, As, Sb). It exhibits either a hexagonal NiAs-type (B81) structure or an orthorhombic MnP-type (B31) structure. In particular, CrAs undergoes a phase transition at 800 K from the NiAs-type to the MnP-type configuration. In the latter phase, the unit-cell lattice parameters are $a$=5.649 {\AA}, $b$=3.463 {\AA} and $c$=6.2084 {\AA}.\cite{wu14} The Cr atoms are situated in the centre of CrAs$_6$ octahedra, surrounded by six nearest-neighbour arsenic atoms, as shown in Fig.~\ref{fig:CrAs}; four of the six Cr-As bonds are inequivalent due to the high anisotropy exhibited by this class of compounds.~\cite{noce16}

\subsection{Band structure and Fermi surface}

\noindent In a tight-binding picture, the real space Hamiltonian describing the system in the MnP-phase is given by
\begin{equation}
\label{eqn:tightbinding}
H=\sum_{i,\sigma}\epsilon_{i}c^+_{i\sigma}c_{i\sigma}-\sum_{i,j,\sigma}t_{ij}(c^+_{i\sigma}c_{j\sigma}+h.c.)\, .
\end{equation}\\

It consists of the diagonal energy terms $\epsilon_{i}$ at each Cr or As lattice site and the hopping terms of electrons with spin $\sigma$ between the $i$ and $j$ sites where the Cr or As ions are located. The hopping amplitudes $t_{ij}$ are given by the following integrals
\[
t_{ij}=\bigl\langle\varphi_{n}({\bf r}-{\bf R}_i)|V({\bf r})|\varphi_{m}({\bf r}-{\bf R}_j)\bigr\rangle,
\]\\
\noindent where $V({\bf r})$ is the hopping potential, $n$ and $m$ are the indexes that run over the dimension of the Fock space, ${\bf R}_i$ is the lattice vector associated with the ion position, and  $\varphi_{n}({\bf r}-{\bf R}_i)$ are the Wannier functions, forming, for all $n$ and all ${\bf R}_i$, a
complete orthogonal set.
\noindent As already pointed out, the primitive cell of the CrAs contains four Cr ions and four As ions. Furthermore, for each Cr ion, the electrons we are considering belong to the $d$-orbitals, while for the As ions the orbitals involved are the $p$-orbitals. This implies that the Hamiltonian $H$ in Eq.~(\ref{eqn:tightbinding}) corresponds to a $32\times32$ matrix. The whole Hamiltonian can be partitioned as in Eq.~(\ref{eqn:matrix}),

\begin{equation}
	\label{eqn:matrix}
H=
\left[
\begin{array}{c|c}
H_{CrCr} & H_{CrAs} \\
\hline
H_{AsCr} & H_{AsAs}
\end{array}
\right]\, ,
\end{equation}
where $H_{CrCr}$ stands for a $20\times20$ matrix that describes the $d$-$d$ hoppings among Cr ions, $H_{AsAs}$ is a $12\times12$ one describing the $p$-$p$ As-ions hoppings, and the two sub-matrices  $H_{CrAs}$ and $H_{AsCr}$ correspond to the $d$-$p$ hoppings from Cr to As ions and vice-versa. 

\noindent As we anticipated, the real space Hamiltonian matrix elements have been set according to the outcome of density functional theory calculations,~\cite{noce16} which have been performed by using the VASP package.~\cite{vasp} In such an approach, the core and the valence electrons have been treated within the projector augmented wave method~\cite{vasp1} and with a cutoff of 400~eV for the plane wave basis. All the calculations have been performed using a 12$\times$16$\times$10 $k$-point grid. For the treatment of exchange-correlation, the local density approximation and the Perdew-Zunger~\cite{Perdew} parametrization of the Ceperly-Alder~\cite{Ceperley} data have been considered. After obtaining the Bloch wave functions, the maximally localized Wannier functions~\cite{Marzari97,Souza01} are constructed using the WANNIER90 code.~\cite{Mostofi08} To extract the Cr 3$d$ and As 4$p$ electronic bands, the Slater-Koster interpolation scheme has been used, in order to determine the real space Hamiltonian matrix elements.~\cite{Mostofi08}. In our analysis, we limit ourselves to consider atomic shells bringing substantial hopping parameters. As a consequence, we include in our calculations the nearest neighbours hopping, the second nearest neighbours hopping along the $x$-direction and the diagonal part of the second nearest neighbours hopping along the $y$ and $z$ direction, with the hopping values ranging from 80 meV to 1 eV.

The derivation of the full energy spectrum of the CrAs involves the resolution of the eigenvalue problem for the 32$\times$32 matrix of Eq.~(\ref{eqn:matrix}).
However, since the As bands are located above and below 2 eV from the Fermi level,~\cite{noce16} one can project out the low-lying As 4$p$ degrees of freedom using the L\"{o}wdin down-folding procedure.~\cite{lowdin50}
This method is based on the partition of a basis of unperturbed eigenstates into two classes, which are related by a perturbative formula  giving  the  influence  of  one  class  of  states  on  the  other. In our case, we use the orthonormal Wannier function basis. For energies around the Fermi level, one can treat the sub-dominat low-lying As degrees of freedom according to the L\"{o}wdin procedure and down-fold the $H_{AsAs}$ matrix of Eq.~(\ref{eqn:matrix}). In this way, the resolution of original eigenvalue problem is mapped to that of a corresponding effective $\widetilde{H}_{CrCr}$, whose rank is 20, where $\widetilde{H}_{CrCr}$ is given by~\cite{andersen95}

\begin{equation}\label{eqn:equationL}
\widetilde{H}_{CrCr}(\varepsilon)={H}_{CrCr}-{H}_{CrAs}\left({H}_{AsAs}-\varepsilon\mathbb{I}\right )^{-1}{H}_{AsCr}\, .
\end{equation}

\begin{figure}
	\centering
	\includegraphics[width=8cm]{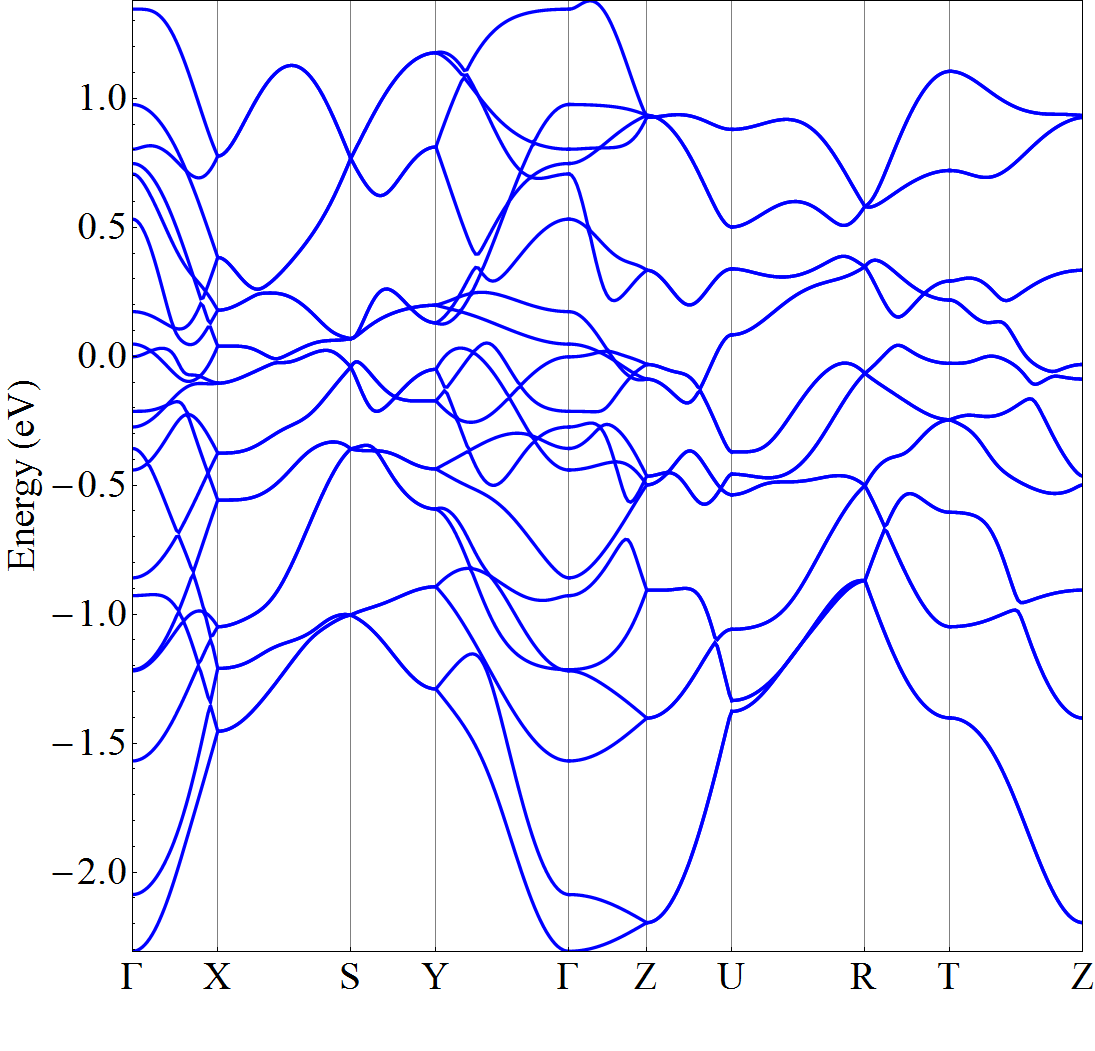}
	\caption{Low energy band structure of the CrAs plotted along high-symmetry paths of the orthorhombic Brillouin zone. The Fermi level is set at zero energy.}
	\label{fig:bandeblu}
\end{figure}

\noindent Using this technique, we get the low energy effective Hamiltonian projected into the Cr-subsector that indeed cannot be obtained using the Wannier function method.

The energy-band dispersions we obtain are shown in Fig.~\ref{fig:bandeblu}, where the high symmetry points along which we plot the band structure have been chosen according to the notation quoted in Ref.~\onlinecite{setyawan10}.
In the non-magnetic phase, the spin up/down channel band structure are degenerate due to the inversion and time reversal symmetries. Moreover, the energy spectrum exhibits an additional degeneracy at the edge of the first Brillouin zone in both spin channels, which is peculiar of the MnP-type orthorhombic configuration.
From an inspection to the figure, we observe that the bandwidth is large almost 3.5~eV; there are flat bands between -1.0 and +1.0 eV and wider bands out of this range due to the hybridization with the As bands. The presence of flat bands around the Fermi level gives rise to van Hove singularities, as may be easily inferred also from an examination of the density of states reported in the next subsection. Furthermore, we can also note a strong anisotropy looking at the band structure along the $\Gamma$-X,
$\Gamma$-Y and $\Gamma$-Z paths.

\begin{figure}
	\centering
	\includegraphics[width=8cm]{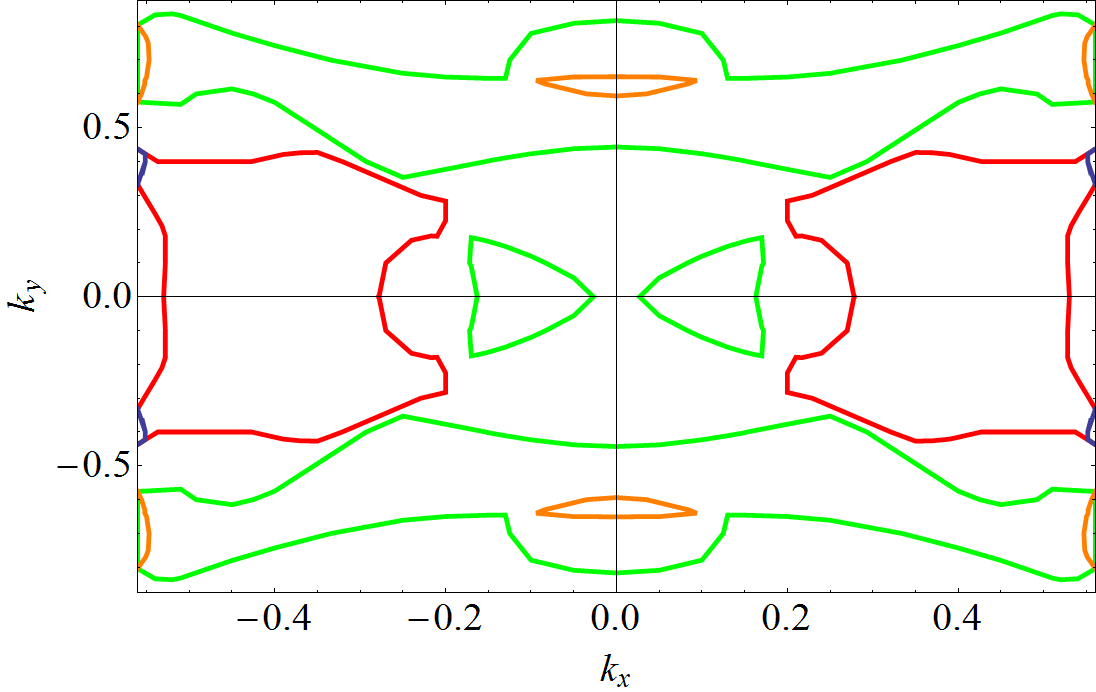}
	\caption{In-plane Fermi surface of CrAs. In-plane ($k_x$, $k_y$) momenta are expressed in units of $1/a$ and $1/b$.}
	\label{fig:fermisurface}
\end{figure}

\noindent Since the energy spectrum displays a little dispersion along the $k_z$ axis, we map in Fig.~\ref{fig:fermisurface} the in-plane Fermi surface. We would like to point-out that we find that the shape as well the pockets of the Fermi surface turn out to be strongly dependent on the magnetic moment of the Cr atoms. In the non-magnetic case, the Fermi surface consists of four sheets while it gets reduced to three when the magnetic moment increases.

\subsection{Density of states}

We use the well known definition of the density of states (DOS):
\begin{equation}
\label{eqn:DOS}
\rho(\epsilon)=\frac{1}{N}\sum_{\boldsymbol{k}}\delta(\epsilon-\epsilon_{\boldsymbol{k}})
\end{equation}
where $\epsilon$ is the energy, $\epsilon_{\boldsymbol{k}}$ is the energy dispersion of the Hamiltonian in Eq.~(\ref{eqn:equationL}) and the sum is carried out on the N values of $\boldsymbol{k}$ in the Brillouin zone.
\noindent The delta functions in Eq.~(\ref{eqn:DOS}) have been approximated by the Gaussian functions:
\begin{equation}
\label{eqn:DOSapprox}
\rho(\epsilon)\approx\frac{1}{N}\sum_{\boldsymbol{k}}\frac{1}{\sigma\sqrt{2\pi}}e^{-\frac{(\epsilon-\epsilon_{\boldsymbol{k}})^{2}}{2\sigma^2}}
\end{equation}
where the expected value of the Gaussians is represented by the eigenvalues of the matrix Eq.~(3) and the variance is assumed to be $\sigma=$ 0.012~eV.\\

\begin{figure}
	\centering
	\includegraphics[width=8cm]{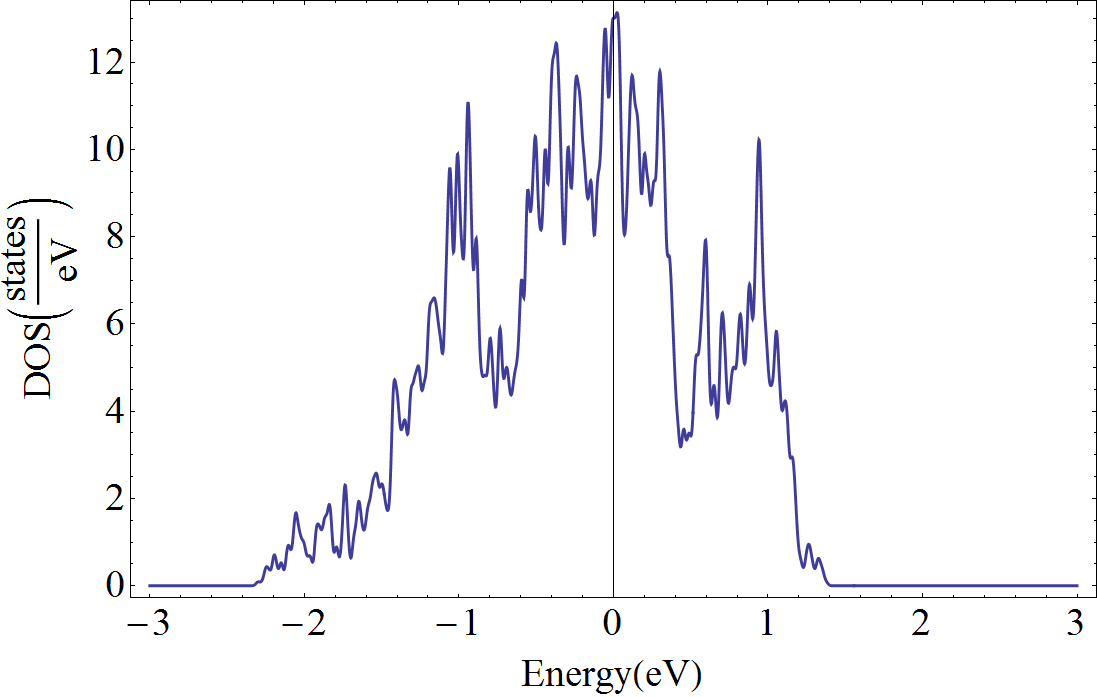}
	\caption{Density of states of the low energy bands of CrAs. The Fermi level is set at zero energy.}
	\label{fig:denstot}
\end{figure}

From an estimation of the spectral weight of the full Hamiltonian in Eq.~(\ref{eqn:equationL}) and from a comparison with the mentioned LDA calculations,~\cite{noce16} we infer that, at the Fermi level, the DOS is predominantly due to Cr electrons, which carry more than 90$\%$ of the total DOS, with a negligible As
contribution. The calculation shows a modest charge transfer from the Cr atoms to the As atoms, which is estimated around 0.4 electrons per atom, again suggesting that the As spectral weight contribution to the low energy states is very small.

Using Eq.~(\ref{eqn:DOSapprox}), we evaluate the DOS related to the band structure of Fig.~2, which is represented in Fig.~4. The main contribution to the DOS is roughly between -1 eV to +1 eV. The DOS shows its maximum close to the Fermi level, allowing for a magnetic instability even at small values of Coulomb repulsion.
Moreover, it presents two peaks at -1~eV and +1~eV, due to some flat Cr bands exhibited by the band structure.

\section{Transport Properties}
The explicit knowledge of the energy spectrum makes also possible the calculation of some transport properties. In the following, as an application, we will show the results for the temperature dependence of the resistivity and the local magnetic moment.

\subsection{Electric Properties}
\begin{figure}
	\centering
	\includegraphics[width=8cm]{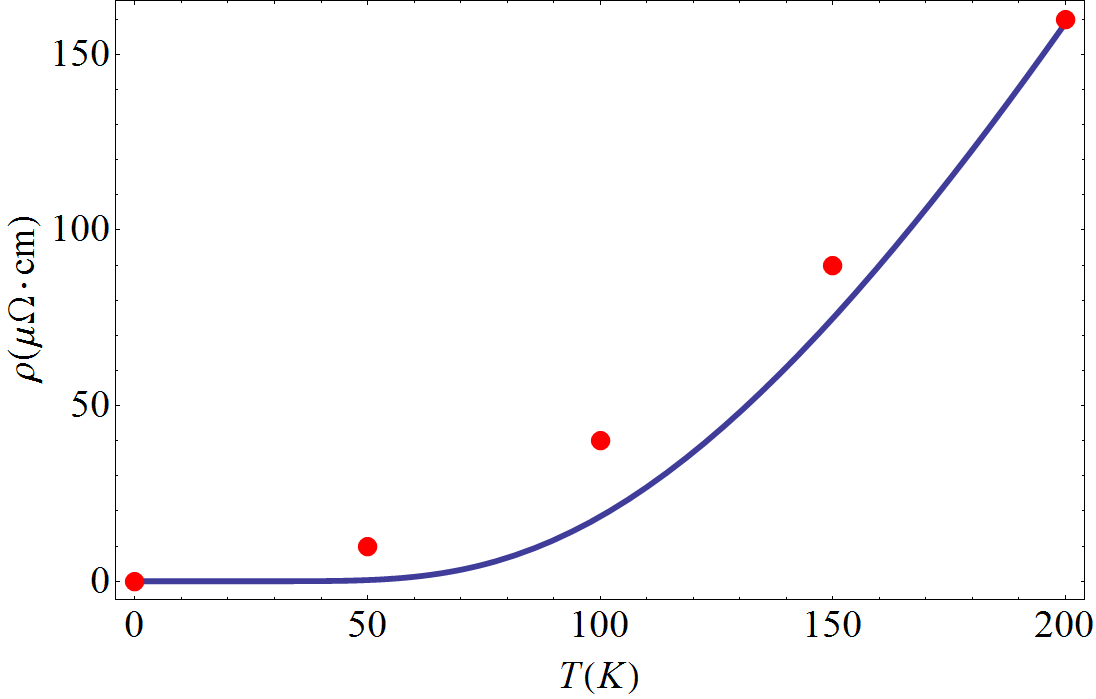}
	\caption{Temperature dependence of in-plane resistivity. The solid line
indicates the outcome of the calculation performed within the Boltzmann theory, while
the circles stand for the experimental data from Ref.~\onlinecite{wu14}.}
	\label{fig:res}
\end{figure}

To evaluate the normal state resistivity for this multi-band case, we calculate the conductivity tensor up to first order.~\cite{ziman79} This calculation implies the knowledge of the energy band spectrum as well as the explicit expression for the relaxation rate.
We start from the following equation for the current density
\begin{equation}
\boldsymbol{J}=e \int \boldsymbol{v} g(\boldsymbol{v}) d\boldsymbol{k}\, ,
\end{equation}
where $e$ and $\boldsymbol{v}$ are the charge and the velocity of electron, respectively, and $g(\boldsymbol{v})$ is the local distribution of electrons. Then, by means of the Boltzmann equation we calculate the local distribution function and finally we get the electron conductivity.~\cite{noce00b} As far as the relaxation times are concerned, we will assume that they depend only on the scattering due to intra and inter-orbital particle-hole correlations, implying a $T^2$ power law.
The result is presented in Fig.~\ref{fig:res} where we suppose, for simplicity, the same relaxation time for all the bands, i. e. $(\tau)^{-1}=\alpha+\beta T^2$, with $\alpha$=1.17 s$^{-1}$ and $\beta$=1.5 x 10$^{-6}$ s$^{-1}$ K$^{-2}$; we recall that  $\tau$ is in units of 10$^{-14}$ second. We find an good overall agreement in a temperature range up to 200~K, where the relaxation time approximation is applicable, i.e., the mean free path is greater than the lattice spacings.

\subsection{Magnetic Properties}

To investigate the magnetic properties within our model, we add to the Hamiltonian in Eq.(1) a local Hubbard term. Then the full Hamiltonian the reads as:

\begin{equation}
\label{eqn:hubbard}
H=-\sum_{<i,j>,\sigma}t_{ij}(c^+_{i\sigma}c_{j\sigma}+h.c.)+U\sum_{i}n_{i\uparrow}n_{i\downarrow},
\end{equation}

\noindent where the last term describes the Coulomb repulsion between electrons, with opposite spin on the same lattice site. Since, as mentioned above, the CrAs is a weakly correlated material, we will treat the previous Hamiltonian within the mean-field approximation:

\begin{equation}
\label{eqn:meanfield}
\begin{split}
H&=-\sum_{<i,j>,\sigma}t_{ij}(c^+_{i\sigma}c_{j\sigma}+h.c.)\\&+U\sum_{i}(n_{i\uparrow}\bigl\langle n_{i\downarrow}\bigr\rangle+n_{i\downarrow}\bigl\langle n_{i\uparrow}\bigr\rangle-\bigl\langle n_{i\uparrow}\bigr\rangle\bigl\langle n_{i\downarrow}\bigr\rangle)
\end{split}\, ,
\end{equation}

\noindent where $\bigl\langle n_{i\uparrow}\bigr\rangle$ and $\bigl\langle n_{i\downarrow}\bigr\rangle$ are the average values of the number operator for spin up and down electrons at $i$ lattice site, respectively.

\begin{figure}
	\centering
	\includegraphics[width=8cm]{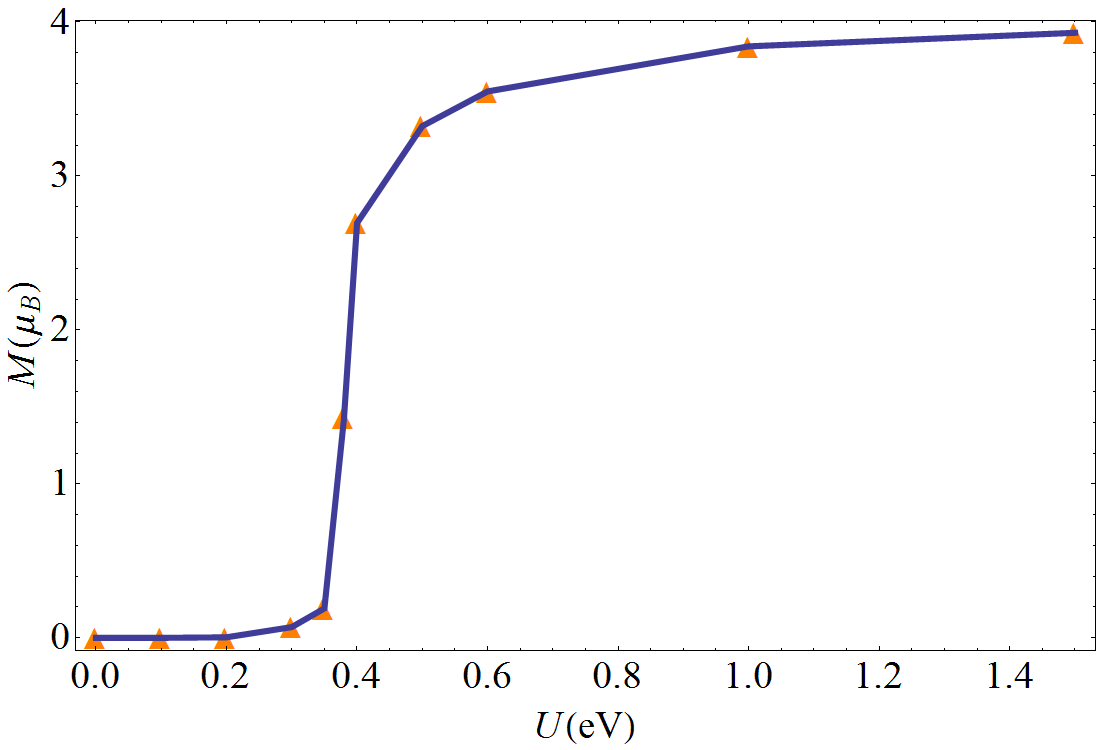}
	\caption{Magnetization of the system evaluated for the mean field Hamiltonian in Eq.~\ref{eqn:meanfield}, as a function of the Coulomb interaction. $U$ is measured in eV  while the magnetization in Bohr magneton units.}
	\label{fig:magnetizz}
\end{figure}

According to LDA calculations\cite{noce16}, there are two different characteristic Cr-Cr distances along the $c$-axis, namely 3.090 and 4.042 {\AA}, corresponding to Cr$_2$-Cr$_3$ and Cr$_1$-Cr$_4$ distances of Cr atoms in Fig.~\ref{fig:CrAs}, respectively. The magnetic coupling between the Cr atoms with shorter distance is strongly antiferromagnetic (-60 meV), while the coupling between Cr atoms with longer distance is weakly ferromagnetic (+10 meV). Moreover, the magnetic ground state is a G-type antiferromagnetic state, with antiparallel nearest neighbours spin.~\cite{noce16}

To obtain the magnetization in the G-type order, we performed a self-consistent procedure according to the following scheme: we assign initial conditions for $n_{i\uparrow}$ and $n_{i\downarrow}$, and use them to evaluate improved expectation values; the procedure runs until convergence is achieved, with the requested accuracy.
In Fig.~\ref{fig:magnetizz} we report the magnetization as a function of the Coulomb interaction $U$. We point out that the Coulomb repulsion on the Cr atoms is around 6~eV in the insulating systems\cite{Autieri2014}. In the metallic phase, the Fermi screening is expected to reduce the electrostatic repulsion so we that we have assumed $U$ varying in the region between 0 and 1.5~eV (orange triangles in Fig.~\ref{fig:magnetizz}).
We can distinguish among three different regimes. The system stays in the non-magnetic phase up $U\approx0.3$~eV. Above this value, one enters a crossover region characterized by unsaturated magnetization which grows as a function of $U$. Above $\approx0.8$~eV, the system is in a fully antiferromagnetic state, with its saturation value 4~$\mu_B$.
It's worth pointing out that the experimental data indicate that the CrAs, in its normal phase, is a metallic itinerant antiferromagnet with a very sensitive magnetic moment to the cell volume and to the magnetic configuration adopted.~\cite{shen16} Our results in Fig.~\ref{fig:magnetizz} suggest that the magnetization strongly depends on the value of Coulomb repulsion in the crossover region, as observed experimentally. We speculate that in that regime 0.3eV$\lesssim U\lesssim0.8$eV, our outcomes show both qualitative and quantitative agreement, the magnetization value being consistent with the experimental estimate of 1.73 $\mu_B$.~\cite{shen16,keller15}

\section{Conclusions}

We have used a combined tight-binding-L\"{o}wdin down-folding approach to calculate the energy bands, the Fermi surface and the DOS of the CrAs. We have also evaluated some magnetic and electric transport quantities finding a good qualitative agreement with the available experimental data.
The analytical formula for low energy bands here presented can be readily used to analyse physical quantities where the topology of the Fermi surface is important as well as the possibility to study the  superconducting instability within the standard broken-symmetry Hartree-Fock scheme.

We would like also to stress that, even if ab-initio calculations are available, they are rather complicated and are not delivered in a form useful as the single-particle term of a model Hamiltonian of Eq.~(\ref{eqn:equationL}), that describes the low-energy excitations. Hence, neglecting the non trivial details of the ab-initio band structure, we have here considered the simplest possible tight-binding model where the most relevant hopping amplitudes are included and where the Hamiltonian projection on the $p$ As subspace has been down-folded according to the L\"{o}wdin procedure. A more sophisticated model starting from the full tight-binding model Hamiltonian of  Eq.~(\ref{eqn:tightbinding}), as well as a different temperature law for the relaxation rates is currently being developed. Calculations in this direction are in progress and will be presented in a forthcoming publication.

\end{document}